\documentclass[
 aip,
 jmp,%
 amsmath,amssymb,
reprint,%
author-year,%
author-numerical,%
]{revtex4-2}

\usepackage{graphicx}% Include figure files
\usepackage{float}
\usepackage{dcolumn}% Align table columns on decimal point
\usepackage{bm}% bold math
\usepackage[utf8]{inputenc}
\usepackage[T1]{fontenc}
\usepackage{mathptmx}
\usepackage{etoolbox}
\usepackage{color}

\begin{document}

%\preprint{AIP/123-QED}

\title{Reversible canted persistent spin textures in two-dimensional ferroelectric bilayer WTe$_{2}$}% Force line breaks with \\
%\thanks{Footnote to title of article.}

\author{Moh. Adhib Ulil Absor}
\email[]{adib@ugm.ac.id}%
\affiliation{Department of Physics, Universitas Gadjah Mada, Sekip Utara BLS 21 Yogyakarta 55281, Indonesia.}

\author{Iman Santoso}
\affiliation{Department of Physics, Universitas Gadjah Mada, Sekip Utara BLS 21 Yogyakarta 55281, Indonesia.}
\date{\today}% It is always \today, today,

\begin{abstract}
The recent discovery of materials hosting persistent spin texture (PST) opens an avenue for the realization of energy-saving spintronics since they support an extraordinarily long spin lifetime. However, the stability of the PST is sensitively affected by symmetry breaking of the crystal induced by external perturbation such as the electric field. In this paper, through first-principles calculations supplemented by symmetry analysis, we report the emergence of the robust and stable PST with large spin splitting in the two-dimensional (2D) ferroelectric bilayer WTe$_{2}$. Due to the low symmetry of the crystal ($C_{s}$ point group), we observe a canted PST in the spin-split bands around the Fermi level displaying a unidirectional spin configuration tilted along the $yz$ plane in the first Brillouin zone. Such a typical PST can be effectively reversed by out-of-plane ferroelectric switching induced by interlayer sliding along the in-plane direction. We further demonstrated that the reversible PST is realized by the application of an out-of-plane external electric field. Thus, our findings uncover the possibility of an electrically tunable PST in 2D materials, offering a promising platform for highly efficient and non-volatile spintronic devices.
\end{abstract}

\pacs{Valid PACS appear here}% PACS, the Physics and Astronomy
\keywords{Suggested keywords}%Use showkeys class option if keyword
\maketitle

\section{INTRODUCTION}

The next generation of spintronics relies on the effective control and manipulation of an electron’s spin degree of freedom without an additional external magnetic field, which is achievable by utilizing the effect of spin-orbit coupling (SOC)\cite{Manchon}. Many intriguing SOC-related phenomena were discovered, including spin relaxation \cite{Fabian, Averkiev}, spin Hall effect \cite{Qi}, spin galvanic effect \cite{Ganichev}, and spin ballistic transport \cite{Lu}. In non-magnetic systems lacking inversion symmetry, the SOC induces momentum $k$-dependent spin-orbit field (SOF) that lifts Kramer’s spin degeneracy and results in chiral spin polarization characterized by the spin-momentum-locking property as manifested by the Dresselhaus \cite{Dress} and Rashba \cite{Rashba} effects. The Dresselhaus effect is generally observed on a system holding bulk inversion asymmetries such as bulk zinc blende \cite{Dress} and wurtzite \cite{Wang_Dress} semiconductors. On the other hand, the Rashba effect is associated with heterostructures and surfaces due to two-dimensional (2D) structural inversion asymmetry, as widely observed on semiconductor quantum-well (QW) \cite{Nitta, Caviglia}, surface heavy metal \cite{LaShell, Koroteev}, and several 2D-layered compounds \cite{Zhuang, Popovi, Absor_R, Affandi, Absor_Pol}. 

Recently, ferroelectric Rashba materials (FRMs) have stimulated the interest of researchers due to the reversible spontaneous polarization and the inherent Rashba effect \cite{Picozzi, Wang2020, Varotto2021}. Here, reversible Rashba spin textures can be achieved in a non-volatile way by switching the direction of the ferroelectric polarization \cite{Picozzi}, which is promising for electrically controllable spintronic devices \cite{PMatos, Wang2020, Varotto2021}. $\alpha$-GeTe was the first FRM predicted \cite{DiSante}, and its reversible spin texture by switching ferroelectric polarization has been experimentally confirmed \cite{Liebmann, Rinaldi}. After that, a handful of ferroelectric materials ranging from similar metal chalcogenides (SnTe) to metal-organic halide perovskites, such as (FA)SnI$_{3}$ \cite{Stroppa}, hexagonal semiconductors (LiZnSb) \cite{Narayan}, and perovskite oxides (HfO$_{2}$ \cite{TaoLL}, and BiAlO$_{3}$ \cite{daSilveira}, were proposed to be FRMs. Despite the robust ferroelectricity, the strong SOC in these materials makes them promising for the realization of FRM-based spintronics devices. However, the strong SOC may induce fast spin decoherence through the Dyakonov-Perel (DP) spin relaxation mechanism \cite{Dyakonov}, reducing spin lifetime, and hence limiting the performance of potential spintronic devices.

A possible way to circumstance this obstacle is to engineer a structure where the SOF orientation is momentum-independent. In a such condition, the spin configuration in the momentum space becomes unidirectional, leading to the well-known persistent spin texture (PST) as was firstly proposed by Schliemann et al. \cite{Schliemann}. The PST  enables a route to overcome spin dephasing and provides non-dissipative spin transport \cite{Bernevig, Altmann}, thus holding great promise for future spintronic applications. Previously, the PST has been demonstrated on semiconductor quantum-well (QW) having equal strength to the Dresselhaus and Rashba SOC \cite{Bernevig, Walser2012, Sasaki2014} or on [110]-oriented semiconductor QWs in which the SOC is described by the [110] Dresselhaus model \cite{Bernevig}. Recently, a more robust PST has been proposed imposing the symmetry of the crystal as reported on bulk BiInO$_{3}$ ($Pna2_{1}$ space group) \cite{Tao2018}. The symmetry-protected PST with purely cubic spin splitting has also been recently reported in bulk materials crystallizing in the $\bar{6}m2$ and $\bar{6}$ point groups \cite{HZhao}. Furthermore, the canted PST driven by the lower symmetry of the crystal has also been reported on ZnO (10$\bar{1}$0) surface \cite{Absor2015} and monolayer transition metal dichalcogenides (TMDCs) including monolayers WTe$_{2}$ \cite{JGarcia} and MoTe$_{2}$ \cite{Vila}. 

Despite these advances, 2D ferroelectric materials with in-plane electric polarization have also been proposed to support the PST \cite{Ai2019, Absor2021c, Absor2019a, Anshory2020, Absor2019b, Lee2020, Absor2021a, Absor2021b, Jia}.  Here, the SOC leads to the unidirectional out-of-plane SOF, resulting in the PST with a unidirectional out-of-plane spin configuration. Various 2D ferroelectric materials including WO$_{2}$Cl$_{2}$ \cite{Ai2019}, Ga$XY$($X$=Se, Te; $Y$=Cl, Br, I) \cite{Absor2021a, Absor2021b}, hybrid perovskite benzyl ammonium lead-halide \cite {Jia}, and group-IV monochalcogenide \cite{Absor2021c, Absor2019a, Anshory2020, Absor2019b, Lee2020}, has been reported to exhibit the PST. However, considering the nature of the in-plane ferroelectricity in these 2D materials, the presence of the PST can be destroyed by disturbing the in-plane ferroelectricity, for an instant, through the application of an external out-of-plane electric field \cite{JagodaIOP, Absor2021c}, thus hindering their practical spintronics. Therefore, finding novel 2D ferroelectric materials with robust and stable PST under the external electric field is highly desirable for practical purposes.   

In this paper, through first-principles density-functional theory (DFT) calculations, we report the emergence of robust and stable PST in the 2D ferroelectric bilayer WTe$_{2}$. Recently, it has been reported that the bilayer WTe$_{2}$ is semimetal exhibiting interlayer sliding ferroelectricity \cite{Fei2018, Pankaj, Wang2019, LiuX, YangQing}, thus achieving the stable PST is expected to be useful for non-volatile spintronics. We find that a unidirectional spin configuration is observed in the spin-split bands around the Fermi level, which is tilted along the $yz$ plane in the first Brillouin zone (FBZ), forming a canted PST. Such a peculiar PST is strongly different from the well-known PST discovered on semiconductor QW \cite{Bernevig, Walser2012, Sasaki2014}, bulk systems \cite{Tao2018, HZhao}, and various 2D ferroelectric materials \cite{Ai2019, Absor2021c, Absor2019a, Anshory2020, Absor2019b, Lee2020, Absor2021a, Absor2021b, Jia}. Our $\vec{k}\cdot\vec{p}$-based symmetry analysis clarified that the observed canted PST in the present systems is enforced by the out-of-plane mirror $M_{yz}$ symmetry of the $C_{s}$ point group. More importantly, we found that the orientation of the canted PST can be effectively reversed upon ferroelectric switching, which is demonstrated through the application of an out-of-plane external electric field. This reversible canted PST, which is not found in the previously reported canted PST on the monolayer TMDCs such as monolayers WTe$_{2}$ \cite{JGarcia} and MoTe$_{2}$ \cite{Vila}, is promising for non-volatile and efficient spintronic applications.

\section{Computational details}

We have performed first-principles DFT calculations based on norm-conserving pseudo-potentials and optimized pseudo-atomic localized basis functions implemented in the OpenMX code \cite{Ozaki, Ozakikinoa, Ozakikinoa}. The exchange-correlation functional was treated within generalized gradient approximation by Perdew, Burke, and Ernzerhof (GGA-PBE) \cite{gga_pbe, Kohn}. The basis functions were expanded by a linear combination of multiple pseudo atomic orbitals (PAOs) generated using a confinement scheme \cite{Ozaki, Ozakikino}, where three $s$-, three $p$-, two $d$-character numerical PAOs were used. The accuracy of the basis functions, as well as pseudo-potentials we used, were carefully bench-marked by the delta gauge method \cite{Lejaeghere}.  

We used a periodic slab to model the bilayer WTe$_{2}$, where a sufficiently large vacuum layer (25 \AA) was applied to avoid the spurious interaction between slabs. We used a $12\times10\times1$ $k$-point and real space grids corresponding to energy cutoffs larger than 300 Ry to obtain the converged results of the self-consistent field (SCF) loops. The minimum-energy pathways of ferroelectric transitions were evaluated by using nudged elastic band (NEB) method based on the interatomic forces and total energy calculations \cite{NEB}. We adopted the modern theory of polarization based on the Berry phase (BP) method \cite{berry} to calculate the ferroelectric polarization. We considered a uniform external electric field perpendicular to the 2D plane of the crystal modeled by a sawtooth waveform during the SCF calculation and geometry optimization. Here, the energy convergence criterion was set to $10^{-9}$ eV. The lattice and positions of the atoms were optimized until the Hellmann-Feynman force components acting on each atom were less than 1 meV/\AA.  

To evaluate the spin-splitting-related properties, we perfomed non-collinear DFT calculations, where the SOC was included self consistently in all calculations by using $j$-dependent pseudo potentials \citep{Theurich}. We calculated the spin textures by deducing the spin vector components ($S_{x}$, $S_{y}$, $S_{z}$) in the reciprocal lattice vector $\vec{k}$ from the spin density matrix. Here, by using the spinor Bloch wave function, $\Psi^{\sigma}_{\mu}(\vec{r},\vec{k})$, obtained from the OpenMX calculations after the SCF is achieved, we calculate the spin density matrix, $P_{\sigma \sigma^{'}}(\vec{k},\mu)$, through the following relation \cite{Kotaka},  
\begin{equation}
\begin{aligned}
\label{1}
P_{\sigma \sigma^{'}}(\vec{k},\mu)=\int \Psi^{\sigma}_{\mu}(\vec{r},\vec{k})\Psi^{\sigma^{'}}_{\mu}(\vec{r},\vec{k}) d\vec{r}\\
                                  = \sum_{n}\sum_{i,j}[c^{*}_{\sigma\mu i}c_{\sigma^{'}\mu j}S_{i,j}]e^{i\vec{R}_{n}\cdot\vec{k}},\\
\end{aligned}
\end{equation}
where $S_{ij}$ is the overlap integral of the $i$-th and $j$-th localized orbitals, $c_{\sigma\mu i(j)}$ is expansion coefficient, $\sigma$ ($\sigma^{'}$) is the spin index ($\uparrow$ or $\downarrow$), $\mu$ is the band index, and $\vec{R}_{n}$ is the $n$-th lattice vector. This method has been successfully applied in our previous studies on various 2D materials \cite{Absor2021a, Absor2021b, Absor2021c, Absor2020, Absor2019a, Absor2019b, Absor2018, Absor2017}

\begin{figure}
	\centering		
	\includegraphics[width=1.0\textwidth]{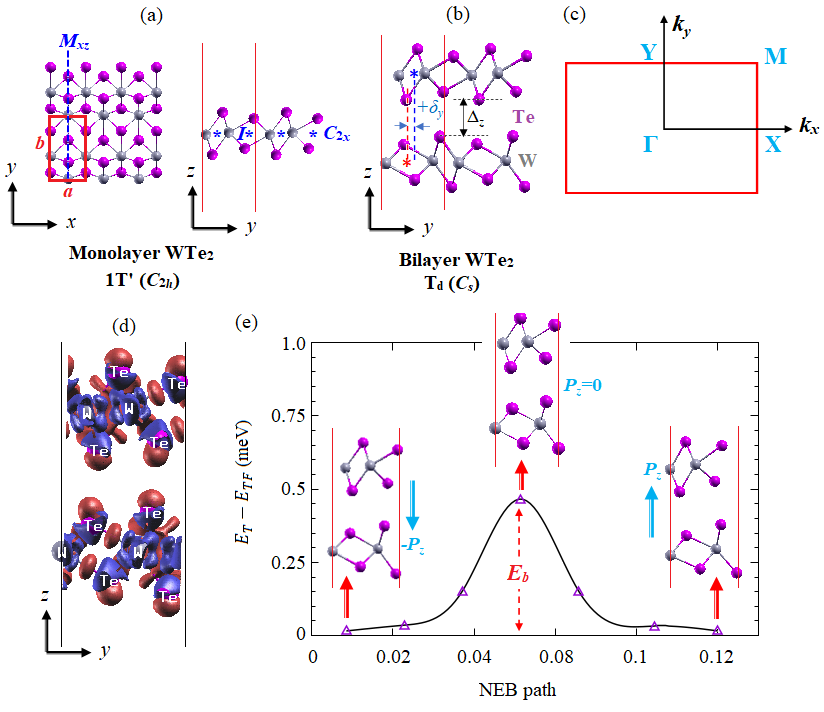}
	\caption{(a)-(b) Atomic structures of monolayer and bilayer WTe$_{2}$, respectively, are presented. The unit cell of the crystal is indicated by the red line and characterized by $a$ and $b$ lattice parameters in the $x$ and $y$ directions, respectively. The symmetry operations of the crystals including inversion symmetry $I$, a mirror symmetry $M_{yz}$ perpendicular to the $x$ axis, and a two-fold screw rotation symmetry $\bar{C}_{2x}$ around the $x$ axis are indicated. $\delta_{y}$ is the in-plane shift along the $y$ direction between the upper and lower layers concerning the inversion center as indicated by the dashed red and blue lines, while $\Delta_{z}$ indicates the interlayer out-of-plane interlayer distance, respectively. (c) The first Brillouin zone of monolayer and bilayer WTe$_{2}$ is shown, where high symmetry $\vec{k}$ points ($\Gamma$, $X$, $Y$, and $M$) are shown. (d) The interlayer differential charge density of bilayer WTe$_{2}$ where red and blue isosurface indicates electron accumulation and depletion after layer stacking, respectively. (e) Minimum-energy pathway of the ferroelectric (FE) transition in bilayer WTe$_{2}$ calculated using Nudged elastic band (NEB) method is presented. $E_{b}$ is the barrier energy defined as the difference between the total energy of the system ($E_{T}$) with respect to the total energy of the ferroelectric ($E_{TF}$). Two FE structures in the ground state with opposite directions of the out-of-plane electric polarization ($P_{z}$) and a paraelectric (PE) structure are inserted. }
	\label{figure:Figure1}
\end{figure}

\section{Results and Discussion}

\subsection{Atomic structure, symmetry, and ferroelectricity}

Figs. 1(a)-(b) show the crystal structures of monolayer and bilayer WTe$_{2}$, respectively, corresponding to the FBZ [Fig. 1(c). Monolayer WTe$_{2}$ is stable in the $1T'$ structure \cite{Hsu, Tang2017, Wang2019}, where the tungsten (W) atoms are octahedrally coordinated by the telluride (Te) atoms and the W atoms form a slightly buckled zigzag chain due to the metallic bonding, resulting in a distortion of the Te octahedron around each Te atom [Fig. 1(a)]. Symmetrically, monolayer WTe$_{2}$ is centrosymmetric belonging to $C_{2h}$ point group\cite{Hsu, Wang2019} generated by the following symmetry operations: (i) identity operation $E$, (ii) inversion symmetry $I$, (iii) a mirror symmetry $M_{yz}$ perpendicular to the $x$ axis, and (iv) a two-fold screw rotation symmetry $\bar{C}_{2x}$ around the $x$ axis [Fig. 1(a)]. Upon van der Walls stacking, bilayer WTe$_{2}$ forms a layered orthorhombic structure known as the $T_{d}$ phase)\cite{Wang2019}, where both the inversion $I$ and screw rotation $\bar{C}_{2x}$ symmetries are broken [Fig. 1(b)]. Therefore, the crystal symmetry of bilayer WTe$_{2}$ becomes $C_{s}$ point group \cite{Wang2019}. We find that the optimized structure of bilayer WTe$_{2}$ for $a$ and $b$ lattice parameters are 3.48 \AA\ and 6.27 \AA, respectively, which are slightly different from that of monolayer WTe$_{2}$ [$a=3.49$ \AA\ and $b=6.31$ \AA]. However, these values are in good agreement with the previous theoretical reports \cite{Wang2019, LiuX, YangQing} and experiments \cite{Fei2018, Pankaj}. Due to the substantial difference between the $a$ and $b$ parameters, the crystal geometry of the monolayer and bilayer WTe$_{2}$ is strongly anisotropic, implying that these materials have a strongly anisotropic mechanical response being subjected to the uniaxial strain along the $x$- and $y$-direction similar to that observed on group IV monochalcogenide \cite{Anshory2020, Kong2018, Liu2019}.   

\begin{figure}
	\centering
		\includegraphics[width=1.0\textwidth]{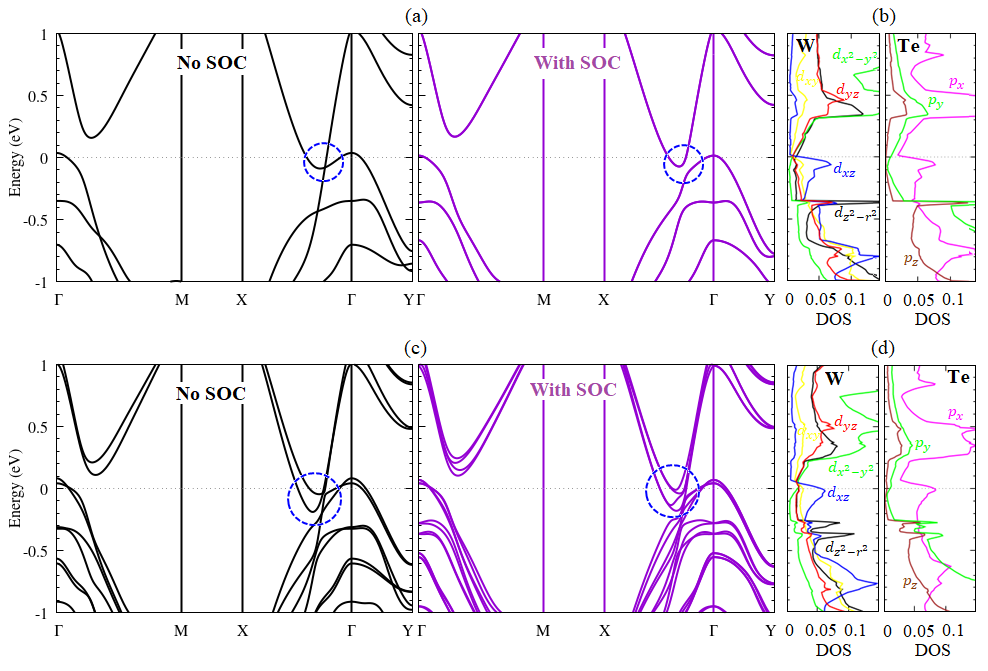}
	\caption{Band structures calculated without (left) and with (right) the SOC corresponding to the density of states projected to the atomic orbitals calculated for: (a-b) monolayer and (c-d) bilayer WTe$_{2}$. The dashed blue lines highlighted the location of the highest occupied state (HOS) and lowest unoccupied state (LUS) near the Fermi level along the $X-\Gamma$ line. In Figs. (b) and (d), each orbital ($p$ and $d$ orbitals) is indicated by the different colors.}
	\label{fig:Figure2}
\end{figure}

The low symmetry of the bilayer WTe$_{2}$ plays an important role in generating the out-of-plane ferroelectricity. Here, the absence of both the inversion $I$ and glide in-plane mirror $\bar{M}_{xy}$ symmetries in bilayer WTe$_{2}$ induces a small in-plane shift ($\Delta_{y}=2\delta_{y}$) between adjacent layers along the $y$ axis (known as in-plane interlayer sliding) \cite{Wang2019, LiuX, YangQing, Fei2018, Pankaj}, where $\delta_{y}$ is the in-plane shift along the $y$-direction between the upper and lower layers with respect to the inversion center as shown by the dashed red and blue lines in Fig. 1(b). In our calculation, the in-plane shift $\Delta_{y}$ is found to be 0.51 \AA, which is much smaller than that of the out-of-plane interlayer distance ($\Delta_{z}$) of 3.19 \AA, but is agree-well with the previously calculated results \cite{Wang2019, LiuX, YangQing}. Accordingly, a net charge transfer between the upper and lower layer is allowed, which is expected to produce an interface dipole in the out-of-plane direction. This is, in fact, confirmed by our calculated results of differential charge density shown in Fig. 1(d), displaying a vertical polarity induced by the electron depletion (blue color) and accumulation (red color) at the interlayer interface.  

To further clarify the robustness of the out-of-plane ferroelectricity in bilayer WTe$_{2}$, we show in Fig. 1(e) the calculated results of the ferroelectric transition pathways obtained from the NEB calculation. We find that two opposite ferroelectric (FE) states with the opposite orientation of the out-of-plane electric polarization ($P_{z}$) are observed, which can be reversed by the in-plane shift ($\Delta_{y}$) between adjacent layers, passing through an intermediate paraelectric (PE) state with $P_{z}=0$ [see the insert of Fig. 1(e)]. The intermediate PE state in bilayer WTe$_{2}$ has a $C_{2v}$ point group due to an additional in-plane glide mirror symmetry $\bar{M}_{xy}$ operation, an in-plane mirror symmetry operation $M_{xy}$ followed by a translation along $y$ by a fractional translation of $1/2a$, thus enforcing the vanishing of $P_{z}$. Through BP calculation, we found that the calculated $P_{z}$ in the FE state is $3.3\times 10^{11}$ $e$ cm$^{-2}$, which is in good agreement with the previous experimental value \cite{Fei2018, Pankaj} and the calculated results \cite{Wang2019, LiuX, YangQing}. Moreover, the transition pathway connecting the two opposite FE states through the PE state yields barrier energy of 0.45 meV [Fig. 1(e)], which is much smaller than that observed on the well-known ferroelectric BaTiO$_{3}$ \cite{Haeni2004}, suggesting the feasibility of a ferroelectric switch under ambient conditions.

\begin{figure}
	\centering
		\includegraphics[width=1.0\textwidth]{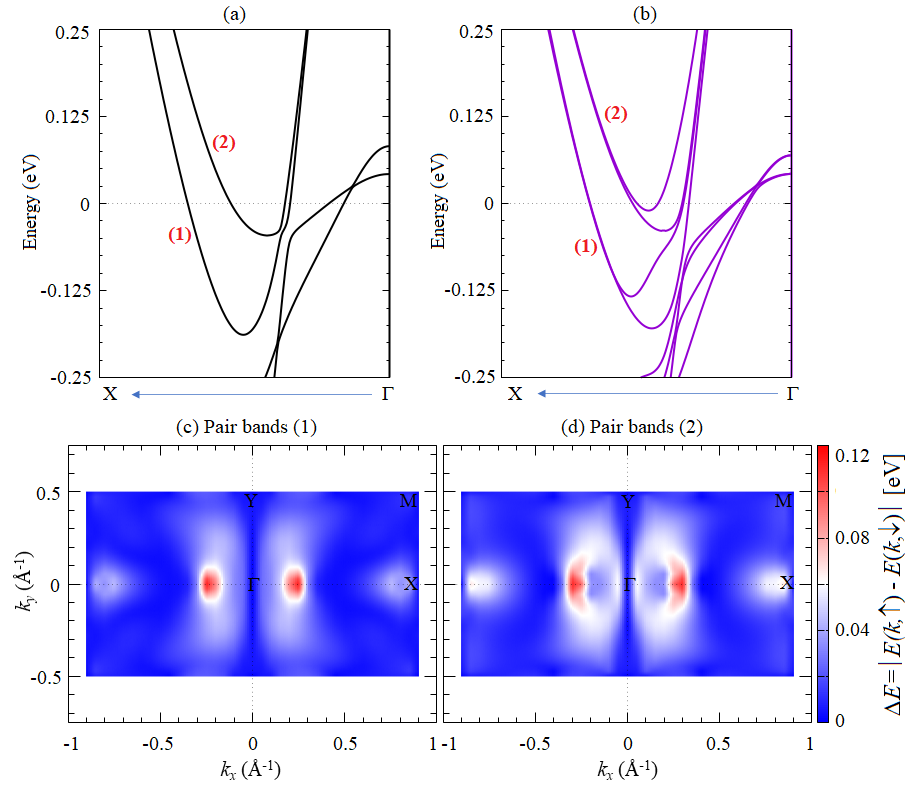}
	\caption{Band structure of bilayer WTe$_{2}$ highlighting the two LUS along the $X-\Gamma$ line (labelled by (1) and (2)) obtained (a(a) without and (b) with the SOC. (c)-(d) Spin-splitting energy map projected to the first Brillouin zone calculated for pair bands (1) and (2) at the LUS, respectively, are shown. The magnitude of the spin-splitting energy, $\Delta E$, defined as $\Delta E=|E(k,\uparrow)-E(k,\downarrow)|$, where $E(k,\uparrow)$ and $E(k,\downarrow)$ are the energy bands with up spin and down spin, respectively, is represented by the color scales.}
	\label{fig:Figure2}
\end{figure}

\subsection{Electronic structure, spin splitting, and spin polarization}

Figs. 2(a)-(c) show the electronic band structure of monolayer and bilayer WTe$_{2}$, respectively, calculated along the selected $\vec{k}$ path in the FBZ. One can see that both the monolayer and bilayer WTe$_{2}$ show a semi-metallic character of the electronic state.  In the absence of the SOC, the monolayer WTe$_{2}$ exhibits band crossing between the highest occupied state (HOS) and lowest unoccupied state (LUS) at the $k$ point near the Fermi level along the $X-\Gamma$ line [see blue dashed lines in Fig. 2(a)], forming a Dirac nodal point \cite{Muechler, Ok, Xu2018, Xiaofeng}, which is protected by $\bar{C}_{2x}$ screw rotation symmetry. Turning the SOC opens this Dirac nodal point and the monolayer WTe$_{2}$ becomes a 2D topological insulator \cite{Muechler, Ok, Xiaofeng, Xu2018}. However, due to the presence of the inversion symmetry in the monolayer WTe$_{2}$, all the bands remain degenerated in the entirely FBZ [see right side of Fig. 2(a)]. Our calculated results of the density of states projected to the atomic orbitals confirmed that the states near the Fermi level (HOS and LUS) mainly originated from the strong hybridization between the W-$d_{xz}$ and Te-$p_{x}$ orbitals [Fig. 2(b)]. 

For the case of the bilayer WTe$_{2}$, on the other hand, we observe two HOS and two LUS near the Fermi level along the $X-\Gamma$ line [Fig. 2(c)] characterized by the strong mixing between the W-$d$ and the Te-$p$ states [Fig. 2(d)], which is similar to the HOS and LUS in the monolayer WTe$_{2}$. However, there is no band crossing (Dirac nodal point) observed in the bilayer WTe$_{2}$ [Fig. 2(c)], which is due to the broken of the $\bar{C}_{2x}$ screw rotation symmetry. In addition, due to the lack of the inversion symmetry in the bilayer WTe$_{2}$, the SOC leads to the significant spin-splitting bands, which is particularly pronounced at the LUS near the Fermi level along the $X-\Gamma$ line [see the right side in Fig. 2(c)]. The emergence of the spin-splitting bands in the bilayer WTe$_{2}$ makes them more beneficial for spintronic applications. Therefore, in the following discussion, we will focus only on the spin-splitting-related properties of the bilayer WTe$_{2}$. 

Figs. 3(a)-(b) show the band structure of bilayer WTe$_{2}$ highlighting the two LUS along the $X-\Gamma$ line (labeled by the band (1) and (2)) obtained without and with the SOC, respectively. Since the LUS are close to the Fermi level, it is expected that they play a significant role in the transport properties of the carriers. By comparing Figs. 3(a) and 3(b), it is clearly seen that large spin splitting is observed at the two LUS, where the spin-splitting energy up to 0.09 eV and 0.12 eV is observed on the pair bands (1) and (2), respectively. These values are comparable to that observed on various 2D ferroelectric systems including group IV monochalcogenides (0.02 - 0.3 eV) \cite{Absor2021c, Absor2019a, Anshory2020, Absor2019b} and Ga$XY$($X$=Se, Te; $Y$=Cl, Br, I) family (0.25 eV) \cite{Absor2021b}, which is certainly sufficient to ensure the proper function of spintronic devices operating at room temperature. By mapping the spin-splitting energy of the LUS projected to the FBZ, we observed a strongly anisotropic spin splitting, where the large spin-splitting energy emerges at the bands along the $X-\Gamma$ line, while very small spin-splitting energy appears at the bands along the $\Gamma-Y$ line [Figs. 3(c)-(d)]. This anisotropic nature of the spin-splitting energy observed in the LUS is enforced by the low symmetry of the crystal that is similar to that observed on various 2D group IV monochalcogenides \cite{Absor2021c, Absor2019a, Anshory2020, Absor2019b} and Ga$XY$($X$=Se, Te; $Y$=Cl, Br, I) family \cite{Absor2021b}.

To further characterize the spin-splitting bands at the LUS, we show the calculated result of spin-resolved projected to the bands as displayed in Fig. 4(a). Due to the presence of the $M_{yz}$ mirror plane in the bilayer WTe$_{2}$, the spin polarization has no $x$-component of spin ($S_{x}$), but exhibits a significant $y$- and $z$-componets of spin ($S_{y}$, $S_{z}$), indicating that the spin polarizations are tilted along the $yz$ plane in the FBZ. By calculating the $k$-space spin textures for the upper and lower bands [Figs. 4(c) and 4(d), respectively] in the spin-split pair bands (1) [see Figs. 3(a)-(b)], we observe a unidirectional spin configuration that is visible at the substantially large region in the FBZ. Such a tilting and unidirectional spin configuration in the $k$-space lead to the formation of the canted PST, which differs strongly from the widely reported PST discovered on semiconductor QW \cite{Bernevig, Walser2012, Sasaki2014}, bulk systems \cite{Tao2018, HZhao}, and various 2D ferroelectric materials \cite{Ai2019, Absor2021c, Absor2019a, Anshory2020, Absor2019b, Lee2020, Absor2021a, Absor2021b, Jia}. We noted here that the canted PST has been previously reported on ZnO (10$\bar{1}$0) surface \cite{Absor2015} and monolayer TMDCs such as monolayers WTe$_{2}$ \cite{JGarcia} and MoTe$_{2}$ \cite{Vila}. The observed canted PST in the present system is expected to induce unidirectional canted SOF in the $k$-space, which protects the spin from decoherence and induces an extremely long spin lifetime \cite{Schliemann, Altmann, Bernevig}. In fact, a robust spin diffusion over long distances has been reported in few-layered MoTe$_{2}$ at room temperature \cite{Song2020}, offering a promising platform to realize an efficient spintronics device.

\begin{figure}
	\centering
		\includegraphics[width=1.0\textwidth]{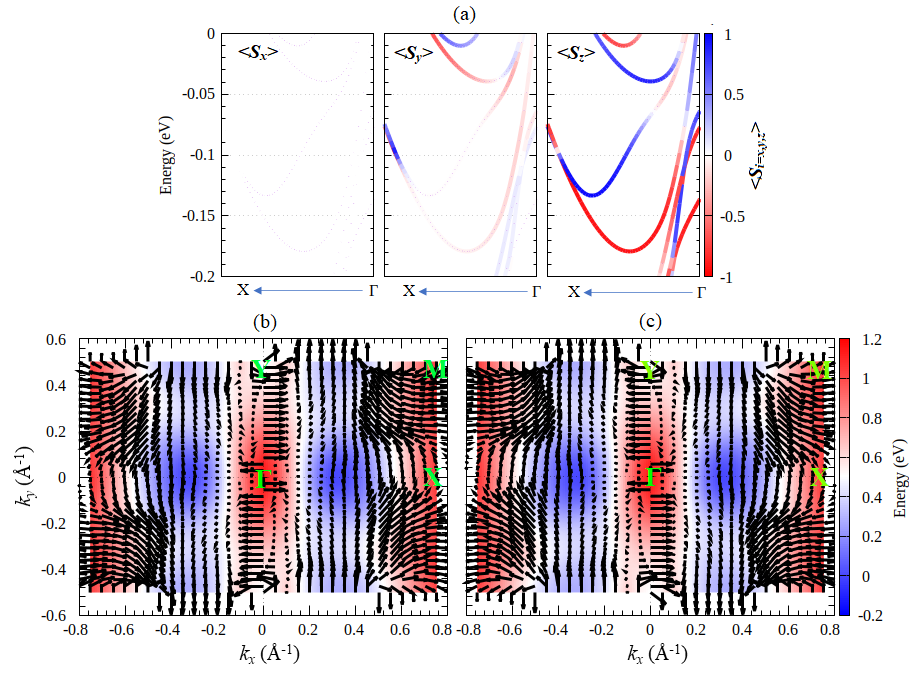}
	\caption{(a) Spin-resolved projected to the bands at the LUS located near the Fermi level along the $X-\Gamma$ line where the expectation value of spin components ($\left\langle S_{x}\right\rangle$, $\left\langle S_{y}\right\rangle$, $\left\langle S_{z}\right\rangle$) represented by the color scale. (b)-(c) Spin texture projected to the FBZ calculated for the upper and lower bands in the pair bands (1) of the LUS, respectively, are shown. The black arrows represent the in-plane spin components of the spin polarization, while the color scale indicates the energy level of the bands.}
	\label{fig:Figure3}
\end{figure}

To understand the nature of the observed canted PST, we consider a symmetry-based $\vec{k}\cdot\vec{p}$ Hamiltonian model recently used for 2D TMDCs with $T_{d}$ structure \cite{JGarcia, Vila}. Since the electronic states near the Fermi level around the $\Gamma$ point belongs to the IRs $\mathcal{A}_{g}$ ($d$ orbital) and $\mathcal{B}_{u}$ ($p$ orbital) of the $C_{s}$ point group, one can construct the following $\vec{k}\cdot\vec{p}$ Hamiltonian:
\begin{equation}
\label{2}
\mathcal{H}_{\Gamma}= \mathcal{H}_{0}+\mathcal{H}_{SOC} 
\end{equation} 
where the first term describes the Hamiltonian without the SOC: 
\begin{equation}
\label{3}
\mathcal{H}_{0}=m_{p} k^{2}\sigma_{0}\tau_{0}+ \left(m_{d}k^{2}+\delta\right)\sigma_{0}\tau_{z}+ \beta k_{y}\sigma_{0}\tau_{y}+ \gamma \sigma_{0} \tau_{x}.
\end{equation}
Here, $\tau_{i}$ ($\sigma_{i}$) with $i=x,y,z$ and $\tau_{0}$ ($\sigma_{0}$) are the Pauli matrices and identity matrix working in the orbital (spin) space, respectively, and $m_{p(d)}$ is the effective mass of the occupied (unoccupied) bands. $\beta$, $\gamma$, and $\delta$ are the parameter representing the degree of the crystalline anisotropy between $k_{x}$ and $k_{y}$, the breaking of the inversion symmetry, and the degree of the band inversion, respectively. The second term in Eq. (\ref{2}) represents the SOC term which can be written in the first order in $k$ as
\begin{equation}
\label{4}
\mathcal{H}_{SOC}= \left(\alpha_{1} k_{x}\sigma_{y}+ \alpha_{2} k_{y}\sigma_{x}+ \alpha_{3} k_{x}\sigma_{z}\right) \tau_{x},
\end{equation}
where $\alpha_{1}$, $\alpha_{2}$, and $\alpha_{3}$ are the SOC parameters. Eq. (\ref{4}) is the symmetry-allowed SOCs obtained by considering the mirror symmetry $M_{yz}=i\sigma_{x}\tau_{0}$ and time reversal symmetry $\mathcal{T}=i\sigma_{y}\mathcal{K}$, where $\mathcal{T}^{2}=-1$ for the spinor and $\mathcal{K}$ is the complex conjugation. 

By focusing on the spin-split states along the $\Gamma-X$ line ($k_{y}=0$), the following simplified Hamiltonian holds,
\begin{equation}
\label{5}
\mathcal{H}_{\Gamma-X}= m_{p} k_{x}^{2}\sigma_{0}\tau_{0}+\left(m_{d}k_{x}^{2} +\delta\right)\sigma_{0}\tau_{z} + \gamma \sigma_{0} \tau_{x}+ \left(\alpha_{1} k_{x}\sigma_{y}+ \alpha_{3} k_{x}\sigma_{z}\right)\tau_{x}.
\end{equation} 
From the last two terms of Eq. (\ref{5}), it is obvious that the spin textures should be unidirectionally tilted along the $yz$ plane, which is agree-well with the observed spin-resolved bands shown in Fig. 4(a) and the $k$-space spin texture presented in Figs. 4(b)-(c). Moreover, the SOC parameters ($\alpha_{1}$, $\alpha_{3}$), which are important for spintronics device operations, can be evaluated by fitting the energy dispersion obtained from the solution of Eq. (\ref{5}) to the DFT energy bands along the $\Gamma-X$. For the lowest unoccupied bands [see the pair bands (1) in Fig. 3(b)], we find that the calculated SOC parameters are 0.09 eV\AA\ and 1.15 eV\AA\ for $\alpha_{1}$ and $\alpha_{3}$, respectively. These values are much larger than that observed on various semiconductor QW \cite{Walser2012, Sasaki2014}.

It is noted here that the observed canted PST in the spin-split bands of the LUS near the Fermi level makes them possible to be resolved by using spin-polarized angle-resolved photo-electron spectroscopy (SARPES). Recently, Fanciulli et.al successfully measure the spin polarization of electrons near the Fermi level in the bulk WTe$_{2}$ by combining a SARPES measurement with a high-harmonic generation laser source \cite{Fanciulli}. Therefore, considering the similar electronic state, resolving the canted PST in the bilayer WTe$_{2}$ is plausible. Moreover, it is also feasible to observe the formation of the PST by injecting the electron in the LUS, which could be performed by using near-field scanning Kerr microscopy \cite{Rudge} to resolve the features down to tens-nm scale with sub-ns time resolution. Furthermore, exploration of the current-induced spin polarization driven by the Edelstein effect \cite{Edelstein} and associated spin-orbit torques \cite{Gambardella} is also possible in the bilayer WTe$_{2}$ by applying an electron doping. Since the electron doping on the bulk WTe$_{2}$ has been experimentally realized \cite{Li_2022}, observation of the current-induced spin polarization in the bilayer WTe$_{2}$ under the electron doping is plausible, which is expected to show the better resolution due to the stronger SOC.

\subsection{Reversible Canted PST}

To enrich the physics and possible application of the present systems, we examine the correlation between the observed canted PST and ferroelectricity. Here, an interesting property known as reversible canted PST holds, i.e., the orientation of the unidirectional tilted spin configuration is reversed by switching the direction of the out-of-plane spontaneous electric polarization $P_{z}$. Fig. 5 displays the evolution of the spin-resolved bands near the Fermi level along the $X-\Gamma$ line in the bilayer WTe$_{2}$ under the different orientation of the out-of-plane ferroelectric polarization $P_{z}$. Interestingly, when the direction of $P_{z}$ is reversed, the out-of-plane spin component ($S_{z}$) is invariant, while the in-plane spin component $S_{y}$ is reversed simultaneously [Figs. 5(a) and 5(c)]. As a result, a reversible canted PST is achieved as schematically shown in Figs. 5(b) and 5(d). Such a reversible PST offers a promising platform to realize a non-volatile control of the PST through the application of the external electric field, which is useful for spintronics devices implementing spin Hall effect \cite{PMatos, Wang2020, Varotto2021}.

\begin{figure}
	\centering
		\includegraphics[width=1.0\textwidth]{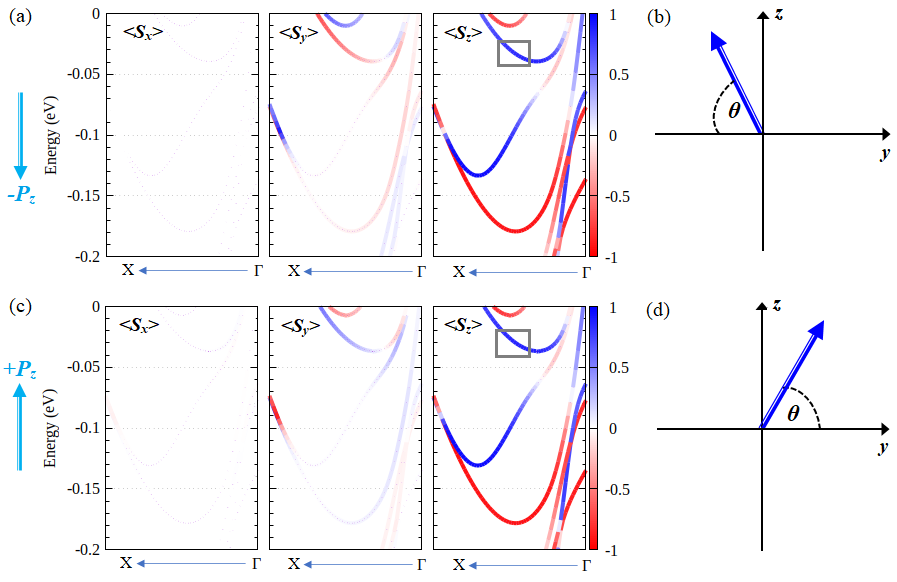}
	\caption{Relation between the out-of-plane ferroelectric polarization $P_{z}$ and the spin polarization. (a) Spin-resolved projected bands for the bilayer WTe$_{2}$ having $-P_{z}$ orientation of the electric polarization and (b) the schematic view of the canted PST are shown. The canted PST is evaluated for the spin polarization at a certain $k$ as indicated by the black line. (c)-(d) Same with Figs. 5(a)-(b) but for the bilayer WTe$_{2}$ with $+P_{z}$ orientation of the electric polarization. The expectation value of spin components ($\left\langle S_{x}\right\rangle$, $\left\langle S_{y}\right\rangle$, $\left\langle S_{z}\right\rangle$) are represented by the color scale.}
	\label{fig:Figure4}
\end{figure}

The physical mechanism behind the reversible canted PST can be understood in terms of symmetry analysis. Let us considered a general symmetry operator $g=\left\{\Omega|\vec{t}_{R}\right\}$, where $\Omega$ is point group symmetry operation and $\vec{t}_{R}$ is the translation vector operator. Since the spin polarization vector $\vec{S}(\vec{k})$ is a time-reversal pseudovector, for a given symmetry operator $g$, the $\vec{S}(\vec{k})$ can be transformed by the relation $\vec{S}^{'}(\vec{k})=g\vec{S}(\vec{k})=\mathcal{P}_{R}\mathcal{P}_{t}\vec{S}(\vec{k})$, where $\mathcal{P}_{R(t)}$ is the spatial (temporal) parity operator correlated with $g$. In particular, $\mathcal{P}_{t}=\pm 1$ when $\hat{\Omega}(\vec{k})=\pm\vec{k}+\vec{G}$, where $\vec{G}$ is the multiples of the reciprocal lattice vector. As for the bilayer WTe$_{2}$, the reversal polarization ($P_{z}$) direction between the two FE states is related to the PE state having glide in-plane mirror symmetry $\bar{M}_{xy}$ [see Fig. 1(d)], thus we obtained that $\mathcal{P}_{R}=-1$ and $\mathcal{P}_{t}=1$. Accordingly, the reversal canted PST can also be correlated to $\bar{M}_{xy}$ through the following transformation: $\left(S_{y}, S_{z}\right)^{+P_{z}}=\left(-S_{y}, S_{z}\right)^{-P_{z}}$. This shows that the canted PST is reversed as the sign of the $S_{y}$ spin component is reversed, which is consistent-well with the observed spin-resolved bands and the spin textures are shown in Fig. 5. 

We emphasized here that the reversible canted PST observed in the present study is the key for the nonvolatile spintronic applications. Here, the application of an electric field is an effective method to realize the reversible canted PST. Therefore, we introduce an external electric field ($E_{z}$) oriented perpendicular to the 2D surface ($\hat{z}$-direction), which can be implemented through the application of a gate voltage \cite{JagodaIOP, Absor2021c}. Since the reversible canted PST in bilayer WTe$_{2}$ depends on the sign of the $S_{y}$ spin component [see Figs. 5(a) and 5(c)], we further show in Fig. 6(a) the modulation of the $S_{y}$ under the different sign and magnitude of the electric field $E_{z}$. Obviously, we identify the reversible spin polarization as indicated by the reversal sign of the $S_{y}$ under the switching direction of $E_{z}$.

\begin{figure}
	\centering
		\includegraphics[width=1.0\textwidth]{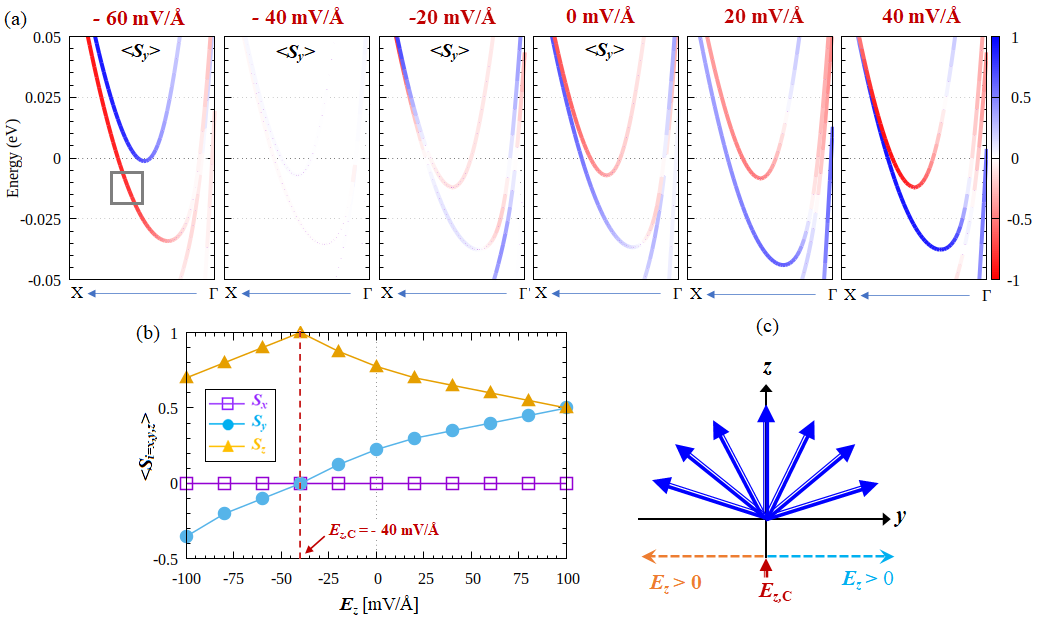}
	\caption{(a) Evolution of the spin-resolved projected bands for the $S_{y}$ spin component under the external out-of-plane electric field $E_{z}$ is shown. (b) The dependence of the expectation value of spin on the electric field $E_{z}$ calculated for the particular $k$ point highlighted by the black line in Fig. 6(a) is presented. The critical electric field $E_{z}^{\texttt{c}}$ showing the paraelectric state is identified. (c) A schematic illustration of the reversible canted PST under the out-of-plane electric field $E_{z}$ is shown.}
	\label{fig:Figure5}
\end{figure}

By calculating the spin polarization at the certain $k$ [see the black line in Fig. 5(a) for the representation], the $E_{z}$-dependent of the expectation value of spin components ($S_{x}$, $S_{y}$, $S_{z}$) are presented in Fig. 6(b). When $E_{z}$ is applied along the $+z$ direction, all the spin components have the same sign as that of the equilibrium system. However, the sign of the $S_{y}$ becomes fully reversed when $E_{z} > 40$ mV/\AA\ is applied along the $-\hat{z}$ direction. At the electric field of -40 mV/\AA, defining the critical field ($E_{z,\texttt{C}}$) applied along the $-\hat{z}$ direction, the PE state of the bilayer WTe$_{2}$ is achieved, thus the $S_{y}$ becomes zero [Fig. 6(b)]. This critical electric field $E_{z,\texttt{C}}$ defines the lower bound for the coercive field in which the orientation of the internal out-of-plane electric polarization $P_{z}$ of the bilayer WTe$_{2}$ starts to flip from the $\hat{z}$ to -$\hat{z}$ direction, which is in fact, one order smaller than that predicted on the 2D ferroelectric group IV monochalcogenide \cite{Hanakata}. The switching sign of the $S_{y}$ under the substantially small electric field $E_{z}$ indicates that an electrically reversible canted PST is achieved; see Fig. 6(c) for schematic representation, thus putting forward the bilayer WTe$_{2}$ as a promising candidate for efficient and nonvolatile spintronic devices.  

Here, we discuss a possible application of the ferroelectric bilayer WTe$_{2}$ in terms of the reversible canted PST. One of the potential applications is using the present system as a barrier in a tunnel junction with a ferromagnetic electrode to observe a tunneling anomalous Hall effect (TAHE) controlled by the switchable ferroelectric polarization \cite{Su_Jing}. Here, tunneling electron through the ferroelectric barrier with a sizable SOC strength leads to an imbalance in the number of the transmitted electron with opposite transverse momenta. Accordingly, a transfer charge current is induced manifesting the TAHE. When ferroelectric polarization of the barrier is reversed, the change sign of the SOC parameters is achieved as dictated by the reversing of the canted PST. As a result, the change sign of the anomalous Hall conductance can be detected, thus enabling a nonvolatile electric field control of the TAHE.  

Next, we highlight the main difference of the reversible canted PST found in the present system with the reversible PST observed on various 2D materials with in-plane ferroelectricity such as WO$_{2}$Cl$_{2}$ \cite{Ai2019}, Ga$XY$($X$ = Se, Te; $Y$ = Cl, Br, I) \cite{Absor2021a, Absor2021b}, hybrid perovskite benzyl ammonium lead-halide \cite{Jia}, and group-IV monochalcogenide \cite{Absor2021c, Absor2019a, Anshory2020, Absor2019b, Lee2020}. In these materials, the fully out-of-plane PST are symmetrically driven by the in-plane mirror (or glide in-plane mirror) $M_{xy}$ symmetry of the $C_{2v}$ point group, which can be reversed by switching the orientation of the in-plane ferroelectric polarization. However, such a reversible PST can be destroyed by breaking the $M_{xy}$ through the application of the electric field perpendicular to the 2D material surface. In fact, the breaking of the PST by the out-of-plane electric field has been previously reported on 2D group IV monochalcogeide \cite{JagodaIOP, Absor2021c}, thus limiting the spintronics functionality. In contrast, for our systems, the reversible canted PST is enforced by the out-of-plane mirror symmetry $M_{yz}$ of the $C_{s}$ point group, which is controllable by the out-of-plane ferroelectric polarization. Since the $M_{yz}$ is invariant under the out-of-plane electric field, the robust and stable PST is achieved [see Figs. 6(a)-(b)], thus the present system is more beneficial for spintronics applications.

Finally, we compare the reversible canted PST found in the present system with the previously reported canted PST on the monolayer TMDCs \cite{JGarcia, Vila}. Similar to our system, the previous works reported the emergence of the canted PST in the monolayers WTe$_{2}$ \cite{JGarcia} and MoTe$_{2}$ \cite{Vila} crystallizing in $T_{d}$ structure. However, because the previous systems are not ferroelectric, the reversible canted PST cannot be achieved, implying that the previous systems cannot be implemented as non-volatile spintronics. In addition, the monolayers WTe$_{2}$ and MoTe$_{2}$ are more stable in the $1T'$ structure rather than the $T_{d}$ structure in the ground state \cite{Hsu, Tang2017, Wang2019}. Since the $1T'$ structure is centrosymmetric ($C_{2h}$ point group), the spin-polarization should be zero due to the degenerated bands in all entirely FBZ [see Fig. 2(a)], thus the canted PST cannot be observed.

\section{Conclussion}

In summary, through first-principles DFT calculations supplemented by symmetry analysis, we have systematically investigated the SOC-related properties of the ferroelectric bilayer WTe$_{2}$, a novel 2D material having a coexistence between the ferroelectricity and semimetal property. We have observed the emergence of the novel type of PST in the spin-split states around the Fermi level, dubbed canted PST. This particular PST exhibits a unidirectional spin configuration tilted along the $yz$ plane in the FBZ, which is significantly different from the previous PST widely observed on various semiconductor QW \cite{Bernevig, Walser2012, Sasaki2014}, bulk systems \cite{Tao2018, HZhao}, and 2D ferroelectric materials \cite{Ai2019, Absor2021c, Absor2019a, Anshory2020, Absor2019b, Lee2020, Absor2021a, Absor2021b, Jia}. Our $\vec{k}\cdot\vec{p}$-based symmetry analysis has clarified that the observed canted PST is enforced by the out-of-plane mirror $M_{yz}$ symmetry of the $C_{s}$ point group in the bilayer WTe$_{2}$. More importantly, we have shown that this typical PST can be effectively reversed upon the out-of-plane ferroelectric switching, which has been demonstrated through the application of an out-of-plane external electric field. Therefore, our study suggested that the present system is promising for efficient and non-volatile spintronics devices. 

Since the canted PST found in the present study is driven by the out-of-plane mirror symmetry $M_{yz}$ operation in the $C_{s}$ point group, it is expected that this typical PST can also be achieved on other 2D materials having the similar point group symmetry. Recently, there have been several other 2D bilayer systems that were predicted to maintain the interlayer sliding ferroelectricity
and low symmetry of the crystals, such as ZrI$_{2}$ \cite{Ding, ZhangT} and VS$_{2}$ \cite{Liu}, thus opening a possibility to further explore the achievable PST in these materials. Our predictions are expected to stimulate further theoretical and experimental research in the exploration of PST-based 2D materials, widening the range of the 2D materials for future spintronic applications.

\begin{acknowledgments}
This work was supported by PD Research Grants (No.1709/UN1/DITLIT/Dit-Lit/PT.01.03/2022) funded by KEMDIKBUD-DIKTI, Republic of Indonesia. The computation in this research was partly performed using the computer facilities at Universitas Gadjah Mada, Republic of Indonesia.
\end{acknowledgments}

\section*{Data Availability Statement}

The data that support the findings of this study are available from the corresponding author upon reasonable request.

\nocite{*}
\bibliography{Reference1}% Produces the bibliography via BibTeX.

%\bibliography{Reference1}
\end{document}